\documentclass[aps,preprint,nofootinbib]{revtex4}%
\usepackage{amsfonts}
\usepackage{amsmath}
\usepackage{amssymb}
\usepackage{float}
\usepackage{graphicx}%
\providecommand{\U}[1]{\protect\rule{.1in}{.1in}}

\begin{document}
\title{Four-dimensional Traversable Wormholes and Bouncing Cosmologies in Vacuum}
\author{$^{(1)}$Andr\'{e}s Anabal\'{o}n and $^{(2)}$Julio Oliva}
\affiliation{$^{(1)}$Departamento de Ciencias, Facultad de Artes Liberales,
Universidad Adolfo Ib\'{a}\~{n}ez, Avenida Padre Hurtado 750, Vi\~{n}a del Mar, Chile.}
\affiliation{$^{(2)}$Departamento de F\'{\i}sica, Universidad de Concepci\'{o}n, Casilla 160-C, Concepci\'{o}n, Chile.}

\begin{abstract}
In this letter we point out the existence of solutions to General Relativity
with a negative cosmological constant in four dimensions, which contain
solitons as well as traversable wormholes. The latter connect two
asymptotically locally AdS$_{4}$ spacetimes. At every constant value of the
radial coordinate the spacetime is a spacelike warped AdS$_{3}$. We compute
the dual energy momentum tensor at each boundary showing that it yields
different results. We also show that these vacuum wormholes can have more than
one throat and that they are indeed traversable by computing the time it takes
for a light signal to go from one boundary to the other, as seen by a geodesic observer. We generalize the wormholes to include
rotation and charge. When the cosmological constant is positive we find a
cosmology that is everywhere regular, has either one or two bounces and that
for late and early times matches the Friedmann-Lema\^{\i}tre-Robertson-Walker
metric with spherical topology.

\end{abstract}
\maketitle
\section*{Introduction}

Traversable wormholes are spacetimes that connect two far away regions by
means of a throat. A major open problem in physic is whether their existence
can take place in physically sensible and simple circumstances. It is
well-established that asymptotically flat gravity in four dimensions requires
exotic matter fields or to go beyond General Relativity to produce a wormhole
\cite{Visser:1995cc}. Holography \cite{Maldacena:1997re}, uses asymptotically
AdS gravity to describe a conformal field theory (and its deformations). In
this context, a careful study shows that Einstein wormholes with a boundary of
positive curvature do not not exist \cite{Witten:1999xp}. From the holographic
point of view it is problematic to define a field theory on a negative
curvature manifold. This is because the scalar fields in the dual field theory
are conformally coupled to the scalar curvature. Therefore, they have an
effective negative squared mass, which would spoil the stability of the
system. We circunvent this dilemma by using AdS itself as the boundary of the
wormhole. Indeed, in AdS, there are tachyonic masses that do not introduce
instabilities and the squared mass of the conformally coupled scalar fields,
although negative, is always safe in this regard \cite{Breitenlohner:1982bm}.
When picking the AdS$_{3}$ boundary it is easy to see that the bulk solution
presented below is a smooth deformation in such a way that the surfaces of
constant radial coordinate are spacelike warped AdS \cite{Bengtsson:2005zj}.
Therefore, it is also natural to consider wormholes where the boundary itself
is warped. This is exactly what is done in this paper.

Another major open problem in physics is to find a non-singular description of
the Big-Bang. A realization of this idea is known as bouncing cosmologies,
which have been shown to be compatible with cosmological data and a viable
alternative for inflation \cite{Novello:2008ra}. However, before this article,
no simple example of a bouncing cosmology was known with no ad-hoc matter
fields or exotic kinetic terms \cite{Cai:2012va}. We use only the Einstein
equations and a positive cosmological constant. The crucial step to construct
this long sought spacetime is to allow for space anisotropies at the bounce.
Notwithstanding these anisotropies we show that, by an adequate election of
the parameters in the metric, the late evolution of the spacetime can be
chosen to be exactly the everywhere homogeneous and isotropic de Sitter
spacetime. Thus, our bouncing cosmologies provide a new arena to explore the
cosmology of our Universe without the problem of the initial singularity.

The mathematics involved in our construction are fairly simple. We review some
of the most interesting geometrical ingredients in the first two sections.
First, we show how to deform AdS$_{3}$ in a smooth way and without introducing
closed timelike curves \cite{Bengtsson:2005zj}. As we shall see our Einstein
wormhole is exactly a spacelike warped AdS spacetime at every constant value
of the radial coordinate. The boundary can be warped or not, depending on an
integration constant that controls the warping at infinity. Later we discuss
how the slicing of AdS$_{4}$ by AdS$_{3}$ superficially resembles a wormhole.
However, the existence of a globally defined change of coordinates from the
AdS$_{3}$ slicing to the sphere slicing can be used to proof that global
AdS$_{4}$ has a single boundary. Then we propose an ansatz to construct a
spacetime with a (warped) AdS$_{3}$ boundary. The wormhole spacetime arises
thus naturally. It is then shown that it can have either a single or two
throats and an anti-throat. The time that takes a photon to go from one
boundary to the other is computed. We then give an elegant argument on the
absence of closed time like curves on the spacetime. We compute the dual
energy momentum tensor at each boundary, yielding different results. The
wormhole solution is then embedded in a general class of metrics that contains
all Einstein black hole solutions in four dimensions. This allow us to obtain
its charged and spinning form. Finally, we make analogous considerations when
the cosmological constant is positive, obtaining a bouncing cosmology that can
be de Sitter at late or early times.

\section*{The Spacelike Warped AdS$_{3}$}

AdS$_{3}$ with radius $\lambda$ can be written as%

\begin{equation}
ds_{AdS_{3}}^{2}=\frac{\lambda^{2}}{4}\left[  -\cosh^{2}\theta dt^{2}%
+d\theta^{2}+\left(  du+\sinh\left(  \theta\right)  dt\right)  ^{2}\right]
\text{,} \label{met}%
\end{equation}
where the coordinates satisfy $\left(  t,\theta,u\right)  \in\mathbb{R}^{3}$.
The isometry of AdS$_{3}$, $SO(2,2)$, is broken to $SL(2,\mathbb{R}%
)\times\mathbb{R}=GL(2,\mathbb{R})$ in the warped metric:
\begin{equation}
ds_{WAdS_{3}}^{2}=\gamma_{ij}dx^{i}dx^{j}=\frac{\lambda^{2}}{\nu^{2}+3}\left[
-\cosh^{2}\theta dt^{2}+d\theta^{2}+\frac{4\nu^{2}}{\nu^{2}+3}\left(
du+\sinh\left(  \theta\right)  dt\right)  ^{2}\right]  \text{ .} \label{ex}%
\end{equation}
The spacetime (\ref{ex}) is a smooth manifold, free of closed timelike curves.
It is a Lorentzian version of the squashed three pseudosphere. Spacelike
warped AdS$_{3}$ \cite{Bengtsson:2005zj}, and their black holes
\cite{Anninos:2008fx}, have been extensively studied and they arise as
solutions of topologically massive gravity with graviton mass $\mu=\frac{3\nu
}{\lambda}$. There is also a pathological version of (\ref{ex}), known as
timelike warped AdS$_{3}$ which does contain closed timelike curves. In this
paper, we shall only focus on the physically relevant case (\ref{ex}).

\section*{Wormhole-like Slicing of AdS$_{4}$}

As is well known, it is possible to slice AdS$_{4}$ in AdS$_{3}$ submanifolds
as follows%
\begin{equation}
g_{\alpha\beta}dx^{\alpha}dx^{\beta}=\frac{\ell^{2}dr^{2}}{r^{2}+1}+\frac
{\ell^{2}}{4}\left(  r^{2}+1\right)  \left[  -\cosh(\theta)^{2}dt^{2}%
+d\theta^{2}+\left(  du+\sinh\left(  \theta\right)  dt\right)  ^{2}\right]
\label{trivial}%
\end{equation}
where $\ell$ is the $AdS_{4}$ radius, i.e.%
\begin{equation}
R_{\alpha\beta}=-\frac{3}{\ell^{2}}g_{\alpha\beta}\text{ ,} \label{EQ}%
\end{equation}
for the metric (\ref{trivial}). While this slicing seems to have a wormhole
throat at $r=0$ and two disconnected boundaries at $r=\pm\infty$, this is just
an artifact of the coordinates. There is a well-known global change of
coordinates that maps (\ref{trivial}) to standard global AdS with a round
sphere at the boundary.

The fact that the two boundaries of (\ref{trivial}) are connected has been
remarked in \cite{Maldacena:2004rf}, where by performing identifications in
the fixed-$r$ manifold it was pointed out that is possible to disconnect the
two boundaries at $r=\pm\infty$. This is simply because the change of
coordinates that maps (\ref{trivial}) to global AdS stop existing as a
by-product of the identification. In the next section we show how is possible
to have a wormhole by resorting to geometry instead of topology.

\section*{The Wormhole Solution}

It is our interest to obtain (\ref{ex}) as the boundary of an asymptotically
AdS$_{4}$ Einstein space. Hence, it is natural to propose the following ansatz%

\begin{equation}
ds^{2}=\frac{4\ell^{4}dr^{2}}{\sigma^{2}f(r)}+g(r)\left(  -\cosh^{2}\left(
\theta\right)  dt^{2}+d\theta^{2}\right)  +f(r)\left(  du+\sinh\left(
\theta\right)  dt\right)  ^{2}\text{ .} \label{metric}%
\end{equation}
The Einstein equations (\ref{EQ}) are satisfied provided%

\begin{equation}
g(r)=\frac{\ell^{2}}{\sigma}\left(  r^{2}+1\right)  \text{ ,}%
\end{equation}

\begin{equation}
f(r)=\frac{4\ell^{2}}{\sigma^{2}}\frac{r^{4}+\left(  6-\sigma\right)
r^{2}+\ell mr+\sigma-3}{r^{2}+1}\text{ ,} \label{f}%
\end{equation}
Where $\sigma$ and $m$ are integration constants. We are interested in the
case where $f$ has no real zero. An straightforward analysis shows that $f$
never vanishes provided%

\begin{equation}
12>\sigma>3\text{ ,}\qquad\left\vert \ell m\right\vert <\frac{2}{3\sqrt{3}%
}\frac{12-\sigma}{\sqrt{\sigma-3}}\text{ .}%
\end{equation}
Thus, for these ranges of the parameters, the metric functions are everywhere
positive and regular and the range of the $r-$coordinate is%

\begin{equation}
\infty>r>-\infty\text{ .}%
\end{equation}
The Kretschmann invariant is%

\begin{align}
R^{\mu\nu\alpha\beta}R_{\mu\nu\alpha\beta}  &  =\frac{24}{\ell^{4}}%
-\frac{12\left(  r^{2}-1\right)  \left(  \left(  r^{2}+1\right)  ^{2}%
-16r^{2}\right)  \left[  4\left(  \sigma-4\right)  ^{2}-m^{2}\ell^{2}\right]
}{\ell^{4}\left(  r^{2}+1\right)  ^{6}}\nonumber\\
&  +\frac{96\left(  \sigma-4\right)  rm\left(  r^{2}-3\right)  \left(
3r^{2}-1\right)  }{\ell^{3}\left(  r^{2}+1\right)  ^{6}}\text{ .}%
\end{align}
It is possible to see that for $\sigma=4$ and $m=0$, the spacetime is
everywhere constant curvature and coincides with (\ref{trivial}). The
interpretation of (\ref{met}) as a wormhole is now straightforward.

As shown by figure 1, the wormhole goes from having a single throat for
$\sigma\leq6$ to have two throats for $\sigma>6$. The two throats must have a
local maximum in between that we call an anti-throat.

\begin{figure}[ptb]
\centering
\includegraphics[width=0.6\textwidth]{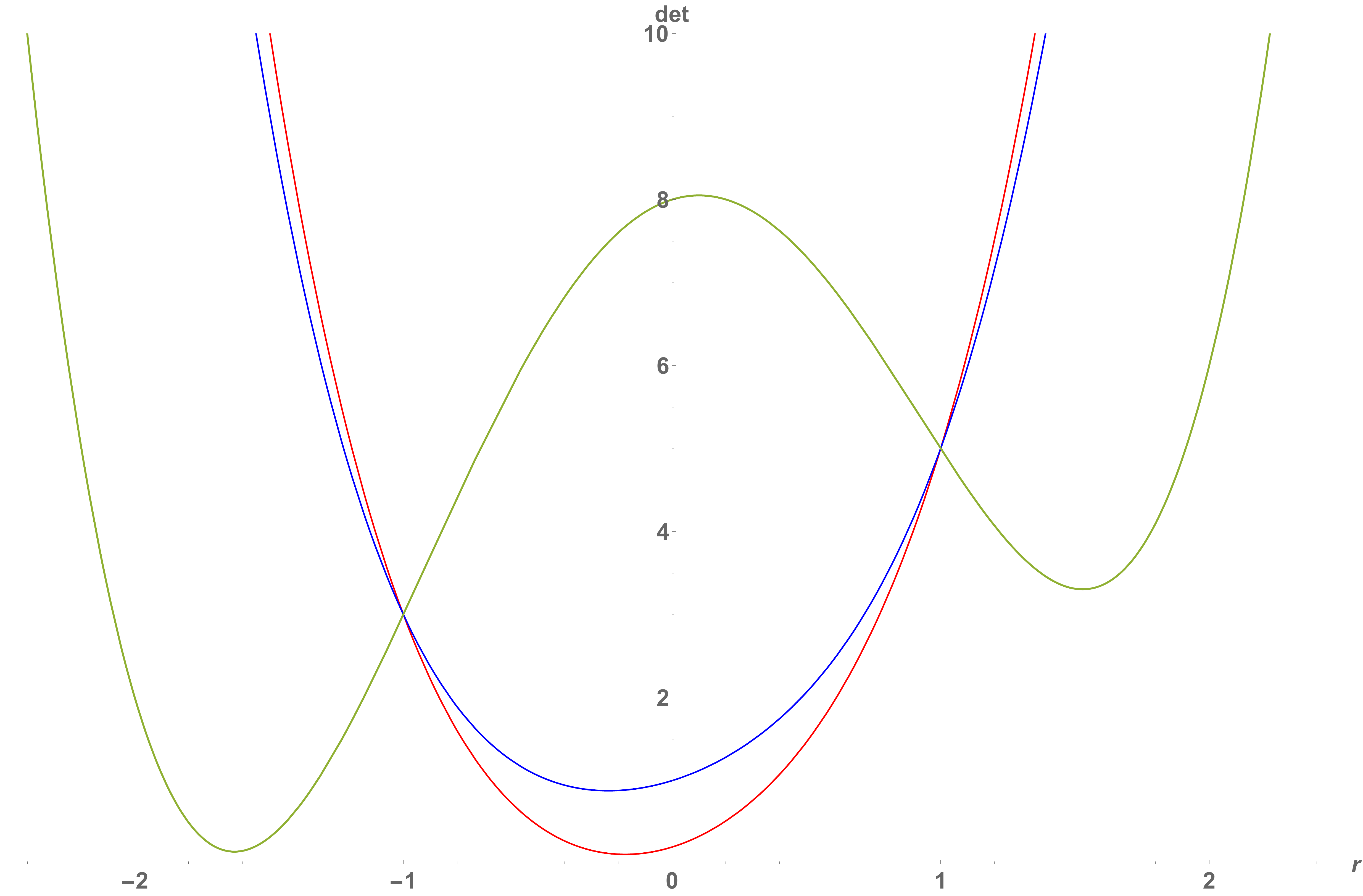} \caption{Here we plot the
dimensionless determinant of the spatial sections with constant, $(t,r)$
$\det=f(r)g(r) \sigma^{3} \ell^{-4}/4$ versus the $r$ coordinate. All the
plots have $m\ell=1$. The plots are for $\sigma=3.2$ and $\sigma=4$ that have
a single throat (from down up) and $\sigma=11$ with two throats and an
anti-throat. As expected, $\det$ is asymmetric unless $m=0$. For a given $m$
all curves intersect at $r=\pm1$ as $f$ is independent of $\sigma$ there.}%
\label{fig1}%
\end{figure}

\bigskip

The scaled timelike coordinate $\ell t/\sigma^{1/2}$ coincides with the proper
time of a geodesic observer located at $r=0=\theta$. According to this observer, the time it takes for a light ray to
go from one boundary to the other is finite, which is expected since the
spacetime is asymptotically AdS$_{4}$ at both asymptotic regions. The crossing
time is given by%
\begin{equation}
\Delta t=\frac{2\ell^{2}}{\sigma}\int_{-\infty}^{+\infty}\frac{dr}%
{\sqrt{f\left(  r\right)  g\left(  r\right)  }}\ .
\end{equation}
To make this integral, and plot it, it is conveniente to write the metric
function $f(r)$ in term of its roots, namely
\begin{equation}
f(r)=\frac{4\ell^{2}}{\sigma^{2}}\frac{\left(  r-z_{1}\right)  \left(
r-z_{1}^{\ast}\right)  \left(  r-z_{2}\right)  \left(  r-z_{2}^{\ast}\right)
}{r^{2}+1}\text{ ,}%
\end{equation}
with $\sigma=2\xi_{1}^{2}+6-\xi_{2}-\zeta$, $z_{1}=\xi_{1}+I\sqrt{\xi_{2}}$,
$z_{2}=-\xi_{1}+I\sqrt{\zeta}$ and $\zeta=\frac{\left(  \xi_{1}^{2}+1\right)
\left(  3-\xi_{1}^{2}-\xi_{2}\right)  }{1+\xi_{1}^{2}+\xi_{2}}$. $f(r)$ has no
real zero provided $\xi_{2}>0$ and $0<\xi_{1}^{2}<3-\xi_{2}$. This is the
region we have plotted. Figure 2 depicts the crossing time as a function of
the parameters $(\xi_{1},\xi_{2})$.
\begin{figure}[ptb]
\centering
\includegraphics[width=0.6\textwidth]{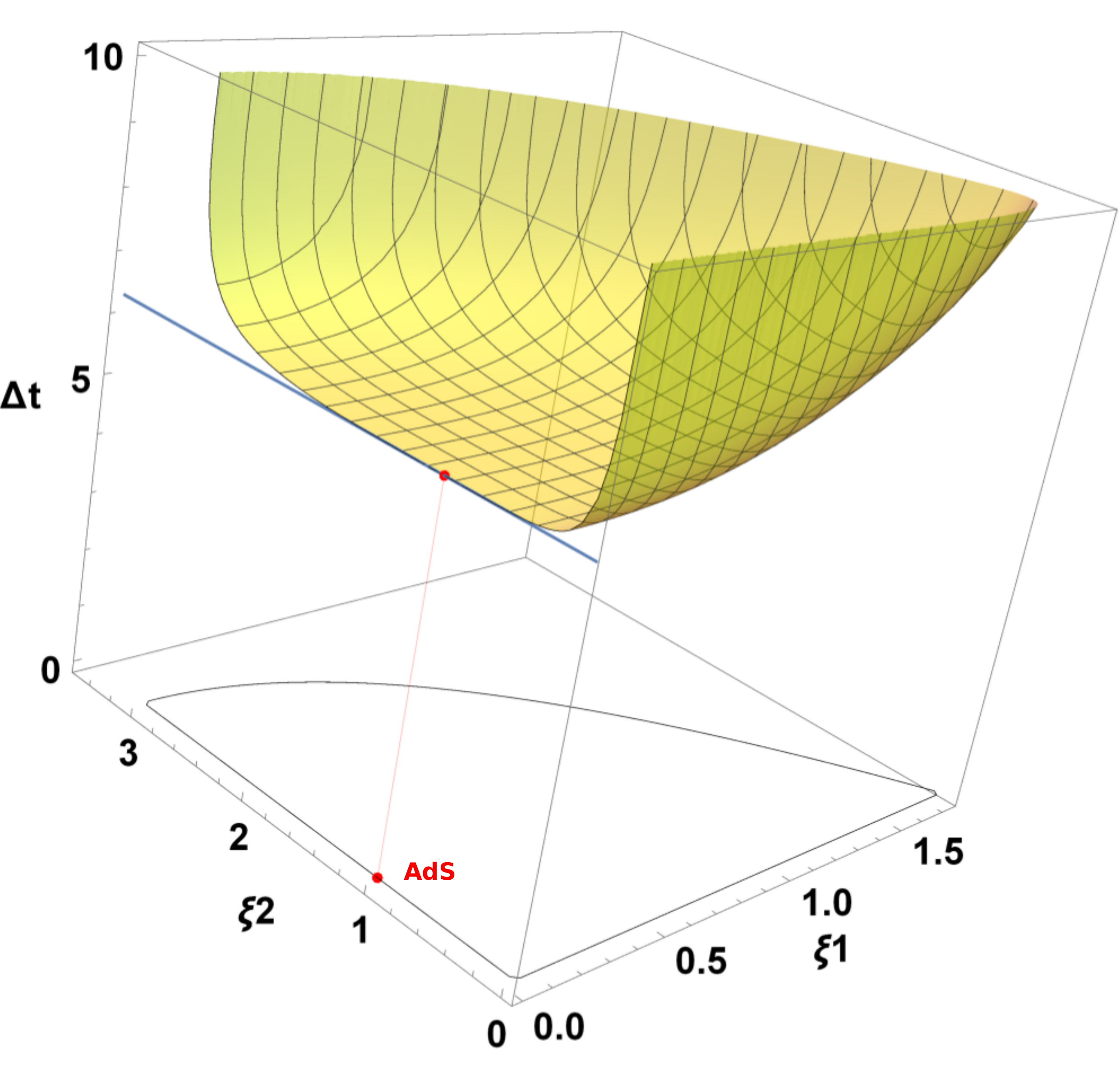} \caption{Crossing time
$\Delta t$ as a function of the parameters $0<\xi_{1}<\sqrt{3}$ and $0<\xi
_{2}<3-\xi_{1}^{2}$. The red dot corresponds to $m=0$ and $\sigma=4$, i.e.
global AdS$_{4}$ which does not represent a wormhole. $\Delta t$ has a global
minimum at that point ($\Delta t_{min}=2\pi$). Outside of the AdS point the
metric has two asymptotic regions and $\Delta t$ is the time it takes to a
photon to go from one asymptotic region to the other, as seen by a geodesic observer at $r=0=\theta$. As suggested in the plot, if one approaches the
boundaries of the domain on the plot $(\xi_{1},\xi_{2})$, $\Delta t$ grows
unboundedly.}%
\label{fig2}%
\end{figure}

\section*{Absence of closed timelike curves}

This argument is a slight generalization of the one in \cite{Coussaert:1994tu}%
. We note that the coordinates $\left(  t,u,\theta,r\right)  $ in
(\ref{metric}) provide a global covering of the manifold. A closed timelike
curve satisfies

\begin{align}
0  &  <\left(  g(r)\cosh\left(  \theta\right)  ^{2}-f(r)\sinh\left(
\theta\right)  ^{2}\right)  \left(  \frac{dt}{d\tau}\right)  ^{2}%
-2f(r)\sinh\left(  \theta\right)  \frac{dt}{d\tau}\frac{du}{d\tau}\nonumber\\
&  -f(r)\left(  \frac{du}{d\tau}\right)  ^{2}-g\left(  r\right)  \left(
\frac{d\theta}{d\tau}\right)  ^{2}-\frac{1}{f(r)}\left(  \frac{dr}{d\tau
}\right)  ^{2}\text{ ,} \label{CTC}%
\end{align}
where $\tau$ yields a good parametrization of the curve. If the curve is
closed, $t$ must come back to its original value. Hence, there must be a point
where $\frac{dt}{d\tau}=0$. Taking into account that $f$ and $g$ are
everywhere positive functions, it is straightforward to see that $\frac
{dt}{d\tau}=0$ is in contradiction with (\ref{CTC}).

\section*{Holographic renormalization}

We now pass to find what is the dual energy momentum tensor associated to this
spacetime. The procedure is as follows. The action, including boundary
counterterms \cite{Balasubramanian:1999re, Mann:1999pc}, is%

\begin{equation}
I[g]=\frac{1}{2\kappa}\int_{M}d^{4}x\sqrt{-g}\left[  R+\frac{6}{\ell^{2}%
}\right]  +\frac{1}{\kappa}\int_{\partial M}d^{3}x\sqrt{-h}\mathcal{K}%
-\frac{1}{\kappa}\int_{\partial M}d^{3}x\sqrt{-h}\left[  \frac{2}{\ell}%
+\frac{\ell}{2}\mathcal{R}\left(  h\right)  \right]  \,,\label{action}%
\end{equation}
where $\kappa=8\pi G$, $\mathcal{K}_{\mu\nu}$ is the extrinsic curvature of
the boundary metric, $h_{\mu\nu}=g_{\mu\nu}-N_{\mu}N_{\nu}$ is the induced
metric on the fixed $r$ hypersurfaces and $\mathcal{R}$ its Ricci curvature.
$N_{\mu}=\delta_{\mu}^{r}\sqrt{g_{rr}}$ is the outward pointing normal, where
we assume that $r>0,$ the case with $r<0$ will be discussed below. Varying the
action gives the energy momentum tensor:
\begin{equation}
\kappa\mathcal{T}_{\mu\nu}=\ell\mathcal{G}_{\mu\nu}\left(  h\right)  -\frac
{2}{\ell}h_{\mu\nu}-\mathcal{K}_{\mu\nu}+h_{\mu\nu}\mathcal{K}\,.
\end{equation}
The boundary metric is $\gamma_{ij}=\lim_{r\longrightarrow\infty}\frac
{1}{r^{2}}h_{ij}$, and is given by (\ref{ex}) with $\frac{\lambda^{2}}{\nu
^{2}+3}=\frac{\ell^{2}}{\sigma}$ and $\nu^{2}=\frac{3}{\sigma-1}$. The dual
enery momentum tensor is%

\begin{equation}
\left\langle \mathcal{T}_{ij}^{+}\right\rangle =\lim_{r\longrightarrow\infty
}r\mathcal{T}_{ij}\text{ ,}%
\end{equation}
which yields\bigskip%
\begin{equation}
\bigskip\left\langle \mathcal{T}_{ij}^{+}\right\rangle dx^{i}dx^{j}=\frac
{\ell^{2}}{\sigma\kappa}\left[  -\frac{m}{2}\left(  -\cosh^{2}\left(
\theta\right)  dt^{2}+d\theta^{2}\right)  +\frac{4m}{\sigma}\left(
du+\sinh\left(  \theta\right)  dt\right)  ^{2}\right]  \text{ .}%
\end{equation}
The wormhole spacetime is invariant under the combined changes $r\rightarrow
-r$ and $m\rightarrow-m$. Hence, the energy-momentum tensor for $r<0$ is the
same than $\mathcal{T}_{ij}^{+}$ changing $m\rightarrow-m$. We note that the
factor in front of the energy momentum tensor can be translated to field
theory variables $\frac{\ell^{2}}{\kappa}=\frac{2^{1/2}}{12\pi}k^{1/2}N^{3/2}$
where we have used the standard holographic dictionary to identify $k$ with
the level and $N$ with the rank of the gauge groups of the ABJ(M) theory, see
for instance \cite{Freedman:2016yue}.

\section*{The Charged and Spinning Generalization}

So far we have studied the simplest case where the wormhole is static. It is
natural to generalize the spacetime to introduce charge and spin. An educated
guess lead us to consider the Plebanski-Demianski \cite{Plebanski:1976gy}
family of spacetimes in four dimensions%

\begin{align}
ds^{2}  &  =\frac{1}{\left(  q-Ap\right)  ^{2}}\left[  -\frac{X(p)}{1+\xi
^{2}q^{2}p^{2}}\left(  \frac{\sigma d\tau}{2}-\xi q^{2}d\phi\right)
^{2}+\frac{Y(q)}{1+\xi^{2}q^{2}p^{2}}\left(  d\phi+\xi p^{2}\frac{\sigma
d\tau}{2}\right)  ^{2}\right. \nonumber\\
&  \left.  +\left(  1+\xi^{2}q^{2}p^{2}\right)  \left(  \frac{dq^{2}}%
{Y(q)}+\frac{dp^{2}}{X(p)}\right)  \right]  ~,
\end{align}
with the gauge field%

\begin{equation}
B=p\frac{Q+P\xi pq}{1+\xi^{2}q^{2}p^{2}}\frac{\sigma d\tau}{2}+q\frac{P-Q\xi
pq}{1+\xi^{2}q^{2}p^{2}}d\phi\text{ .}%
\end{equation}
The Einstein-Maxwell equations%

\begin{equation}
R_{\mu\nu}-\frac{1}{2}g_{\mu\nu}R-\frac{3}{\ell^{2}}g_{\mu\nu}=2\kappa\left(
F_{\mu\sigma}F_{\nu}^{\cdot\sigma}-\frac{1}{4}g_{\mu\nu}F_{\alpha\beta
}F^{\alpha\beta}\right)  \text{ ,}\qquad\nabla^{\mu}F_{\mu\nu}=0\text{ ,}%
\end{equation}
with $F_{\mu\nu}=\partial_{\mu}B_{\nu}-\partial_{\nu}B_{\mu}$, are satisfied provided%

\begin{align}
Y  &  =\ell^{-2}-A^{2}\xi^{-2}\left(  Q_{T}^{2}+y^{4}\right)  -y_{1}%
qA+y_{2}q^{2}+y_{3}q^{3}+y_{4}q^{4}\text{ ,}\\
X  &  =\xi^{-2}\left(  Q_{T}^{2}+y^{4}\right)  +y_{1}p-y_{2}p^{2}-Ay_{3}%
p^{3}+\left(  \xi^{2}\ell^{-2}-A^{2}y_{4}\right)  p^{4}\text{ ,}%
\end{align}
where $Q_{T}^{2}=\kappa Q^{2}+\kappa P^{2}$. This solution is known to contain
all spinning black holes in four dimensions as special limits. To retrieve the
wormhole we found that is necessary to set the acceleration parameter to zero,
$A=0,$ make the following change of coordinates%

\begin{equation}
\tau\left(  t\right)  =2\ell^{2}t\text{ ,}\qquad q\left(  r\right)  =\frac
{1}{r\sigma}\text{ ,}\qquad p(y)=y+\frac{1}{\sigma\xi}\text{ ,}\qquad
\phi(t,u)=2\ell^{2}\left(  u-\frac{t}{K\sigma^{2}\xi}\right)  \text{ ,}%
\end{equation}
and the reparemeterization$\qquad$%

\begin{equation}
y_{2}=\ell^{-2}\sigma^{-2}\left(  6+\epsilon\sigma\right)  \text{ ,}\qquad
y_{4}=\ell^{-2}\sigma^{-4}\left(  y_{0}\xi^{2}\sigma^{3}-\xi N\sigma
^{2}-\epsilon\sigma-3\right)  -Q_{T}^{2}\text{ ,}%
\end{equation}

\begin{equation}
y_{1}=\ell^{-2}\sigma^{-3}\left(  N\sigma^{2}+2\epsilon\sigma\xi
^{-1}+8\right)  \text{ .}%
\end{equation}
Then if $Q=0$, the metric and the gauge field have a smooth $\xi=0$ limit
which exactly coincide with the wormhole with $y=\sinh(\theta)$ (\ref{metric})
provided $\epsilon=-1$, $N=0$, $y_{0}=1$ and $Q_{T}=0$. When $P\neq0$ the
static wormhole is charged. $\epsilon$ controls the topology of the boundary.
For $\epsilon=0$ the boundary has no curvature and when $\epsilon=1$ the
curvature is positive. It can be seen that there are wormholes only when
$\epsilon=-1,$ otherwise the spacetime describe AdS solitons.

If $\xi\neq0$ then the interpretation of the spacetime as a wormhole is less
simple. The metric is singular at $r=0$ and $y=-1\,$. At every constant radial
coordinate there is a black hole, which is regular at the $r=\pm\infty$
boundaries. The black hole flows into the bulk through the $r$ coordinate.
Only at $r=0$ a singularity is developed.

\section*{Bouncing Cosmologies}

The existence of wormholes when the cosmological constant is negative motivate
us to look for bouncing cosmologies when the cosmological constant is
positive. The relevant Einstein metric, $R_{\alpha\beta}=\frac{3}{\ell^{2}%
}g_{\alpha\beta},$ is now%

\[
ds^{2}=-\frac{4\ell^{4}dt^{2}}{\sigma^{2}f(t)}+g(t)\left(  \cos\left(
\theta\right)  d\phi^{2}+d\theta^{2}\right)  +f(t)\left(  d\psi+\sin\left(
\theta\right)  d\phi\right)  ^{2}\text{ .}%
\]
with $g(t)=\frac{l^{2}}{\sigma}(t^{2}+1)$ and%

\begin{equation}
f(t)=\frac{4\ell^{2}}{\sigma^{2}}\frac{t^{4}+(6-\sigma)t^{2}+\mu t+\sigma
-3}{t^{2}+1}%
\end{equation}
with exactly the same form than (\ref{f}). Therefore, the same analysis
applies here regarding regularity. The number of bounces and anti-bounces the
spacetime can have is described by figure 1. The change of coordinates
$t=\exp(\frac{\tau}{\ell})$, when $\sigma=4$ yields for large $\tau$%

\begin{equation}
ds^{2}=-d\tau^{2}+\frac{l^{2}}{4}\exp\left(  \frac{2\tau}{\ell}\right)
\left[  \cos\left(  \theta\right)  d\phi^{2}+d\theta^{2}+\left(  d\psi
+\sin\left(  \theta\right)  d\phi\right)  ^{2}\right]  +O(1)
\end{equation}
which is just the Friedmann-Lema\^{\i}tre-Robertson-Walker metric with
spherical topology. There is a straightforward generalization of this
cosmology along the lines of the previous section.

\section*{Discussion}

In this paper we have constructed the first geometrically non-trivial family
of wormhole solutions to four dimensional Einstein gravity with a negative
cosmological constant. The hypersurfaces perpendicular to the radial
coordinate are warped AdS$_{3}$ spacetimes with the warping that is running
along this coordinate. The asymptotic form of the constant-$r$ metric can be
either warped or not. The wormhole is traversable and free of closed time like
curves. We have generalized the geometry and shown that it is a special limit
of the more general charged Plebanski-Demianski spacetime. These spacetimes
should now be studied at this new light.

Wormhole geometries in four dimensional, asymptotically AdS spacetimes have received large
attention recently due to holography, see for instance \cite{malda2018}. The
holographic dual of a highly entangled state of two non-interacting CFTs is an
eternal black hole, which has two asymptotically AdS regions that are causally
disconnected \cite{Maldacena:2001kr}. It was recently found that the inclusion
of an interaction between the two CFTs opens a throat in the bulk which
causally connects the boundaries \cite{Gao:2016bin}, and the size of the
throat increases with the rotation in the bulk \cite{Caceres:2018ehr}. Our
findings imply that such settings can also take place in vacuum.

When the function $f(r)$ in (\ref{f}) have zeroes, it is possible to cut the
spacetime at the first zero and identify the coordinate $u$ to eliminate the
conical singularity at the degeneration surface of $\partial_{u}$. This
procedure yields a soliton if the zero is of order one. If the zero is of
order two then one simply finds another asymptotic region in the bulk
spacetime. The interior asymptotic region yields in certain cases an RG flow.
The soliton can be thought as a new vacuum of general relativity when the
conformal class of the boundary metrics contain (warped) AdS$_{3}$. These
boundary conditions have been used to holographically describe graphene
\cite{Andrianopoli:2018ymh}.

The introduction of identifications in (\ref{ex}) yields warepd AdS black
holes \cite{Anninos:2008fx}. It is likely that the same identification in
(\ref{metric}) yields black holes together with the flow of the warping
parameter into the radial direction.

The bouncing cosmology presented in the last section yields a smooth
description of the evolution of the Universe. What is remarkable there is that
it is possible to recover an standard homogeneous and isotropic Universe for
late and early times. Nowadays, the experimental data favours a flat universe.
Hence, this cosmological model would only be compatible with the data if the
spherical Universe is large enough. 

\section*{Acknowledgments}

We thank Laura Andrianopoli, Elena Caceres, Bianca Cerchiai and Mario
Trigiante for discussions. This work was supported in part by the Chilean
FONDECYT [Grant No.\ 1181047, 1170279, 1161418], CONYCT-RCUK [Newton-Picarte
Grants DPI20140053 and DPI20140115] and the Alexander von Humboldt foundation.


\end{document}